# Purple bacteria and quantum Fourier transform

Samir Lipovaca
slipovaca@nyse.com

Keywords: purple bacteria, quantum Fourier transform, qubit

*Abstract:* The LH-II of purple bacteria Rhodospirillum (Rs.) molischianum and Rhodopseudomonas (Rps.) acidophila adopts a highly symmetrical ring shape, with a radius of about 7 nm. In the case of Rps. acidophila the ring has a ninefold symmetry axis, and in LH-II from Rs. molischianum the ring has an eightfold symmetry axis. These rings are found to exhibit two bands of excitons. A simplified mathematical description of the exciton states is given in Hu, X. & Schulten, K. (1997) Physics Today 50, 28-34. Using this description, we will show, by suitable labeling of the lowest energy ($Q_y$) excited states of individual BChls, that the resulting exciton states are the quantum Fourier transform of the BChls excited states. For Rs. molischianum ring exciton states will be modeled as the four qubit quantum Fourier transform and the explicit circuit will be derived. Exciton states for Rps. acidophila ring cannot be modeled with an integer number of qubits. Both quantum Fourier transforms are instances of the hidden subgroup problem and this opens up a possibility that both purple bacteria implement an efficient quantum circuit for light harvesting.

## (1) Introduction

*1.1 General overview of photosynthesis*

Photosynthesis is the most important biological process on Earth. The capture of solar energy by photosynthetic organisms, including plants, algae and a variety of types of bacteria, and its conversion into the chemical energy is the primary energy source of almost all the living world. For example, in plants the light-driven reactions convert energy of light into a stable transmembrane potential. The transmembrane potential results from a difference in electrochemical activity of protons across the cell membrane. It has both electrical and chemical contributions. The first being the difference in electrostatic potential from all ions which affect the energy of the protons and the latter coming from difference in the proton concentration. Oxygen ($O_2$) is evolved and ATP and NADPH are formed during the light reactions. The light reactions are series of electron transfer reactions occurring only during the light illumination. In a series of dark reactions, which can occur in both light and darkness, ATP and NADPH are used to reduce $CO_2$ to form glucose and other vital organic products. Both the light and dark reactions take place in chloroplasts. The light reactions [1-8] occur in two closely coupled pigment systems. Light energy is absorbed by a network of antenna pigment - proteins (light harvesting complexes (LHCs)). It is then very efficiently transported through energy transfer to the photochemical reaction center (RC) where energy is converted through a sequence of electron transfer reactions to the transmembrane potential. The light harvesting complexes and reaction centers are also present in photosynthetic bacteria.

Three types of pigments can be present in these systems which act to absorb light or serve as energy transfer sites: (1) bacteriochlorophyll or chlorophyl, (2) carotenoid and (3) bilin (open chain tetrapyrolle) depending on type of organism (a plant, an algae or a bacteria). In general, pigments are noncovalently bound to proteins, forming the pigment-protein complexes. Light harvesting in plants, as well as in bacteria, is carried out by protein-bound chlorophyl-carotenoid aggregates and involves a hierarchical interplay of the pigment's electronic excitations. Very few chlorophylls in the reaction center directly take part in photochemical reactions. Most chlorophylls serve instead as light-harvesting antennas capturing the sunlight and funneling the electronic excitation toward the RC. The pigment organization in which multiple light-harvesting antennas serve the RC seems to have been adopted by all photosynthetic organisms. With it, they can collect light from a broader spectral range and use energy much more efficiently. In particular, light-harvesting antennas enlarge the RC's cross section for capturing sunlight. By feeding the RC with excitation energy, the light-harvesting complexes keep the RC running at an optimal rate. According to this rationale, photosynthesis is initiated by electronic excitation of an aggregate of light-harvesting pigments and by transfer of the excitation to the RC.

*1.2. Purple bacteria*

Purple bacteria are great masters of harvesting light. They developed their virtuosity in a habitat below that of most plant life - that is, at the bottom of ponds or in topsoil, depending on the species. Only light left unharvested by plants penetrates to these depths, mainly at wavelengths of about 500 nm and above 800 nm. To harvest this light efficiently [9] (at about 98% efficiency), the bacteria exploit elegant quantum physics.

Two well studied purple bacteria are Rhodospirillum *(Rs.) molischianum* [10] and *Rhodopseudomonas (Rps.) acidophila* [11]. In these bacteria, the photosynthetic membranes contain two types of light-harvesting complexes: light-harvesting complex I (LH-I) and light-harvesting complex II (LH-II). LH-I surrounds the RC, whereas LH-II is not directly associated with the RC but transfers excitation energy to the RC by way of LH-I. Light harvesting antenna consists of aggregates of bacteriochlorophylls (BChl a) imbedded in protein.

The LH-II antenna of *Rps. acidophila* contains two circular BChl a aggregates, the B800 ring with 9 weakly interacting BChl a's and the B850 ring with 18 BChl a's with strong interaction. The B850 ring is close to the membrane surface while the B800 ring is in the middle of the membrane bilayer. This arrangement puts each ring of pigments in a different chemical environment, so they absorb at different wavelengths (B850 at 850 nm absorption maxima and B800 at 800 nm), extending the spectral range of light harvested. In LH-II of Rs. Molischianum the corresponding numbers are 8 and 16, respectively. Thus, LH-II of both bacteria adopts a highly symmetrical ring shape, whose radius is about 7 nm. In the case of *Rps. acidophila* the ring has a ninefold symmetry axis, and in LH-II from *Rs. molischianum* the ring has an eightfold symmetry axis.

These B850 rings, as the most remarkable structural element, have been found in quantum chemical calculations of their excitations to exhibit two bands of excitons [9]. The band splitting reflects a weakly dimerized from of the aggregate (the BChl - BChl distances alternate slightly along the rings). Most beautiful is the property of excitons that stems from the fact that they extend coherently over the entire BChl ring: Photons are absorbed only by 2 exciton states - namely, the second and the third energetically lowest states, the corresponding exciton states actually being energetically degenerate.

*1.3. Excitons of Circular Aggregate of Bacteriochlorophylls*

The description that follows is based on a simple model presented in [9]. The B850 BChls of LH-II can be represented as circular aggregates of 2N BChls, where N = 8 and N = 9 for LH-II from *Rs. molischianum* and *Rps. acidophila,* respectively. The aggregates exhibit only N-fold symmetry due to a dimerization of the BChls (reflected in their structures) and to interactions between the electronic excitations of the individual BChls. These excitations can be represented as

$$|j\rangle = |Chl_1, Chl_2, ..., Chl^*_j, ..., Chl_{2N}\rangle, \quad j = 1, 2, ..., 2N. \tag{1}$$

Due to the stated interactions, stationary states of the aggregates are coherent superpositions, so-called excitons, of the lowest energy ($Q_y$) excited states of individual BChls. Naturally, the electronic excitations of neighboring BChls interact most strongly, and in this simplified description all other interactions can be neglected. In addition, neglecting the observed dimerization of BChls in the aggregates we assume a 2N-fold symmetry axis. This leads to the following excitonic states

$$|\tilde{n}\rangle = \frac{1}{\sqrt{2N}} \sum_{j=1}^{2N} e^{ijn\pi/N} |j\rangle \tag{2}$$

with energies

$$\varepsilon_n = E_0 + 2V_0 \cos\frac{n\pi}{N}, \quad n = -N+1, -N+2, ..., N \tag{3}$$

$E_0$ represents the $Q_y$ excitation energy of the individual BChls and $V_0$ the interaction energy between neighboring BChls. The state of the lowest energy, for $V_0 > 0$, is

$$\varepsilon = \varepsilon_N = E_0 - 2V_0 \tag{4}$$

It can be shown that only the second and third lowest electronic excitations of the aggregate carry oscillator strength

$$\varepsilon^+_- = E_0 - 2V_0 \cos\frac{\pi}{N} \tag{5}$$

corresponding to n = N-1, -N+1 respectively.

**(2) Methods**

*2.1 Quantum Computers and Light Harvesting Antenna*

Quantum computers process information in a way that preserves quantum coherence. Unlike a classical bit, a quantum bit, or qubit, can be in a superposition of 0 and 1 at once. This quantum mechanical feature allows quantum computers to perform some computations faster than classical computers. For example, quantum computers if built could factor large numbers [12], search data base [13] and simulate quantum systems more rapidly and efficiently [14] than corresponding classical algorithms [15-22]. It has been shown that the time evolution of the wave function of a quantum mechanical many particle system can be simulated precisely and efficiently on a quantum computer [23]. Physicists believe that all aspects of the world around us can ultimately be explained using quantum mechanics. In this sense a light harvesting antenna is a quantum mechanical many particle system composed of a large number of atoms and the time evolution of its wave function could be simulated precisely and efficiently on a quantum computer. This simulation would completely disclose the light harvesting function. In the light of the quantum information and computation it is tempting to imagine LH-II of purple bacteria as a quantum circuit that processes information (harvested photons) from the environment and funnels it to other LH-IIs or to LH-I that surrounds the RC.

*2.2 The quantum Fourier transform*

One of the most useful ways of solving a problem in mathematics or computer science is to transform it into some other problem for which a solution is known. A great discovery of quantum computation has been that some such transformations can be computed much faster on a quantum computer than on a classical computer. One such transformation is the discrete Fourier transform. In the usual mathematical notation [24], the discrete Fourier transform takes as input a vector of complex numbers, $x_0, \ldots, x_{N-1}$ where the length N of the vector is a fixed parameter. It outputs the transformed data, a vector of complex numbers $y_0, \ldots, y_{N-1}$, defined by

$$y_k = \frac{1}{\sqrt{N}} \sum_{j=0}^{N-1} x_j e^{2\pi i j k / N} \qquad (6)$$

The quantum Fourier transform is exactly the same transformation, although the notation is somewhat different. The quantum Fourier transform on an orthonormal basis |0>,...,|N-1> is defined to be a linear operator with the following action on the basis states,

$$|j\rangle \rightarrow \frac{1}{\sqrt{N}} \sum_{k=0}^{N-1} e^{2\pi i j k / N} |k\rangle. \qquad (7)$$

Remarkably, similarity between equations (2) and (7) is obvious. In the following section we will show how to convert (2) into (7) by suitable labeling of the states that represent individual excitations of the BChls in the ring aggregate.

**(3) Results**

*3.1 Excitons of Circular Aggregate of Bacteriochlorophylls and the quantum Fourier transform*

Using a freedom of defining a wave function up to a phase factor, we will relabel states that represent individual excitations of the BChls in the ring aggregate as

$$|l>' = |Ch_0, Ch_1, ..., Ch_l^*, ..., Ch_{2N-1}>, \quad l = 0, 1, ..., 2N-1 \quad (8)$$

such that $e^{il\pi(1-\frac{1}{N})}|l>' = |j>$ and $l = j - 1$. Explicitly, this means

$$|0>' = |1>, \quad e^{i\pi(1-\frac{1}{N})}|1>' = |2>, ..., e^{i\pi(2N-1)(1-\frac{1}{N})}|2N-1>' = |2N>.$$

This allows us to rewrite (2) as

$$|\tilde{n}> = \frac{1}{\sqrt{2N}} \sum_{l=0}^{2N-1} e^{\frac{i(l+1)n\pi}{N}} e^{il\pi(1-\frac{1}{N})} |l>' \quad (9)$$

We can further manipulate (9) as

$$|\tilde{n}> = e^{\frac{in\pi}{N}} \frac{1}{\sqrt{2N}} \sum_{l=0}^{2N-1} e^{\frac{inl\pi}{N}} e^{il\pi(1-\frac{1}{N})} |l>' \quad (10)$$

or

$$e^{-\frac{in\pi}{N}} |\tilde{n}> = \frac{1}{\sqrt{2N}} \sum_{l=0}^{2N-1} e^{\frac{2inl\pi}{2N}} e^{il\pi(1-\frac{2}{2N})} |l>' \quad (11)$$

If we set 2N = M, (11) becomes

$$e^{-\frac{2in\pi}{M}}|n\tilde{\ }> = \frac{1}{\sqrt{M}}\sum_{l=0}^{M-1} e^{\frac{2inl\pi}{M}} e^{il\pi(1-\frac{2}{M})}|l>' \qquad (12)$$

Let $k = n + N - 1 = n + \frac{M}{2} - 1$.

Since $n = -N+1, -N+2, ..., N$ it follows that $k = 0, 1, ..., 2N-1$. Then (12) can be rewritten as

$$e^{-\frac{2i(k-\frac{M}{2}+1)\pi}{M}}|k - \frac{M}{2} + 1\tilde{\ }> = \frac{1}{\sqrt{M}}\sum_{l=0}^{M-1} e^{\frac{2ilk\pi}{M}}|l>' \qquad (13)$$

The right side of (13) is the quantum Fourier transform of the state $|k>'$

$$|k>' \to \frac{1}{\sqrt{M}}\sum_{l=0}^{M-1} e^{\frac{2ilk\pi}{M}}|l>' \qquad (14)$$

while the left side is the excitonic state of the energy $\varepsilon_n$.

*3.2 Product Representation*

We take $M = 2^n$, where $n$ is some integer and the basis $|0>', ..., |2^n - 1>'$ is the computational basis for a $n$ qubit quantum computer. It is useful [24] to write state $|k>'$ using the binary representation $k = k_1 k_2 ... k_n$. More explicitly, $k = k_1 2^{n-1} + k_2 2^{n-2} + ... + k_n 2^0$. It is convenient to adopt the following notation $0.k_l k_{l+1}...k_m$ to represent the binary fraction $\frac{k_l}{2} + \frac{k_{l+1}}{4} + ... + \frac{k_m}{2^{m-l+1}}$. Then with a little elementary algebra the quantum Fourier transform can be given the following product representation:

$$|k_1, ..., k_n> \to \frac{(|0> + e^{2i\pi 0.k_n}|1>)(|0> + e^{2i\pi 0.k_{n-1}k_n}|1>)...(|0> + e^{2i\pi 0.k_1 k_2 ... k_n}|1>)}{2^{\frac{n}{2}}} \qquad (15)$$

where $|k>' \equiv |k_1, ..., k_n>$. This representation allows us to construct an efficient quantum circuit computing the Fourier transform (14). Such circuit is shown in [24]. The circuit implements Hadamard $H$ and $R_p$ gates:

$$H = \frac{1}{\sqrt{2}}\begin{bmatrix} 1 & 1 \\ 1 & -1 \end{bmatrix} \qquad R_p = \begin{bmatrix} 1 & 0 \\ 0 & e^{\frac{2i\pi}{2^p}} \end{bmatrix}$$

where $p = 2,\ldots,n$. Applying $R_n R_{n-1}\ldots R_2 H$ to the first qubit of the state $|k_1 k_2 \ldots k_n >$ results in the state

$$\frac{1}{2^{\frac{1}{2}}}(|0> + e^{2i\pi 0.k_1 k_2 \ldots k_n}|1>)|k_2 \ldots k_n>. \tag{16}$$

Performing a similar procedure ($R_{n-1} R_{n-2} \ldots R_2 H$) on the second qubit yields the state

$$\frac{1}{2^{\frac{2}{2}}}(|0> + e^{2i\pi 0.k_1 k_2 \ldots k_n}|1>)(|0> + e^{2i\pi 0.k_2 \ldots k_n}|1>)|k_3 \ldots k_n>. \tag{17}$$

Continuing in this fashion we will discover that $R_2 H$ need to be applied on the $n-1$ th qubit and only $H$ on the $n$ th qubit. The final state will be

$$\frac{1}{2^{\frac{n}{2}}}(|0> + e^{2i\pi 0.k_1 k_2 \ldots k_n}|1>)(|0> + e^{2i\pi 0.k_2 \ldots k_n}|1>)\ldots(|0> + e^{2i\pi 0.k_n}|1>) \tag{18}$$

The swap operations are then used to reverse the order of the qubits, transforming (18) to (15).

*3.3 Purple bacteria and the product representation*

The B850 BChls of LH-II can be represented as circular aggregates of 2N BChls, where N = 8 and N = 9 for LH-II from *Rs. molischianum* and *Rps. acidophila*, respectively. Applying the product representation for the respective quantum Fourier transform (14) for *Rs. molischianum* we can derive an efficient circuit shown in the Fig. 1., since $M = 2N = 2^4$.

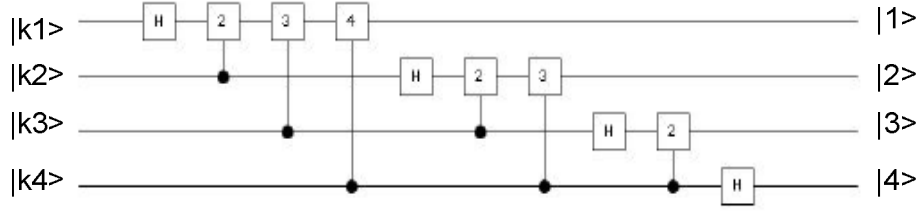

**Fig. 1.** Efficient circuit for the quantum Fourier transform when $n = 4$. $R_p$ gates are labeled as 2, 3, and 4. Not shown are swap gates at the end of the circuit which reverse the order of the qubits.

The output states are, according to the previous section

$$|1> = \frac{1}{\sqrt{2}}(|0> + e^{2i\pi 0.k_1 k_2 k_3 k_4}|1>)$$

$$|2> = \frac{1}{\sqrt{2}}(|0> + e^{2i\pi 0.k_2 k_3 k_4}|1>)$$

$$|3> = \frac{1}{\sqrt{2}}(|0> + e^{2i\pi 0.k_3 k_4}|1>)$$

$$|4> = \frac{1}{\sqrt{2}}(|0> + e^{2i\pi 0.k_4}|1>).$$

and $|1> \otimes |2> \otimes |3> \otimes |4>$ is equal to (18) for $n = 4$. On the other hand, for *Rps. acidophila* $M = 18$ and 18 cannot be expressed as $2^n$ for some integer $n$. Since both circular aggregates of B850 BChls for *Rs. molischianum* and *Rps. acidophila* are almost identical in the structure (2 extra BChls in *Rps. acidophila*) it is natural to express 18 as $2^n$ where

$$n = 1 + 2\frac{\log 3}{\log 2} = 4.169925001.$$

Then the circuit shown in the Fig.1. is a reasonable approximation for *Rps. acidophila* B850 BChls circular aggregate.

**(4) Discussion**

We have shown that the resulting exciton states of the B850 BChls of LH-II from *Rs. molischianum* and *Rps. acidophila* are the quantum Fourier transforms of the BChls excited states. For *Rs. molischianum* ring exciton states can be modeled as the four qubit quantum Fourier transform and we derived an efficient quantum circuit. Exciton states for *Rps. acidophila* are modeled by the same circuit only approximately.

These findings are based on the assumption of neglecting the observed dimerization [9] of BChls in the aggregates as well as taking only into account that the electronic excitations of neighboring BChls interact most strongly. Nevertheless it is tempting to explore a possibility that both purple bacteria implement an efficient quantum circuit. In the case of *Rps. acidophila* it may be noted that such circuit has yet to be found. The circuit shown in the Fig.1. is its reasonable approximation.

It would be too naive to expect observable quantum effects in proteins strongly inhibited by their macroscopic nature and the fact that proteins exist at near room temperature. Although these conditions normally result in a very fast collapse of the respective wave function to one of the allowed classical states it has been suggested [25] that under certain circumstances it is in principle possible to obtain the necessary isolation against thermal losses and other environmental interactions. In such isolation quantum coherence and entanglement, extending over scales that are significantly larger than the atomic scale, may be accomplished and maintained for times comparable to the characteristic times for biomolecular and cellular processes. We would like to argue that these quantum effects are expressed in LH-II of purple bacteria. It is determined that LH-II of *Rs. molischianum* can be considered an aggregate of eight building blocks [9] commonly referred as protomer units. Each protomer unit consists of two helices, three BChls and a carotenoid. Similarly, LH-II of *Rps. acidophila* [11] can be considered an aggregate of nine protomer units each consisting of two helices, three BChls and a carotenoid. BChls and carotenoids in both aggregates harvest the light and the role of the remaining part of the complex (helices) is that of a scaffold to hold the aggregate properly positioned in the membrane of the respective bacterium and to tune absorption maxima for B850 and B800 BChls accordingly. It seems challenging to understand the role of the protein as to provide an isolation of BChls and carotenoids from thermal losses and other environmental interactions maximizing their light harvesting efficiency by allowing, for example, excitons to extend coherently over the entire B850 BChl rings.

By the same token we can expect an entanglement effect expressed in LH-II of purple bacteria. Entanglement is an application of the superposition principle to a composite system consisting of two or more subsystems [26]. A subsystem here is a single BChl molecule. Suppose that BChl 1 can be in one of the two states, A or C, and that these two states represent two contradictory properties, such as being excited and not excited (the ground state). BChl 2, on the other hand can be in one of the two states, B or D. Again these states could represent contradictory properties such as being excited and not excited. The state AB is called a product state. When the entire system is in state AB, we know that BChl 1 is excited and BChl 2 is in the ground state. Similarly, the state CD for the entire system means that BChl 1 is in the ground state and BChl 2 is excited. Now consider the state AB + CD. We obtain this state by applying the superposition principle to the entire, 2 - BChl system. The superposition principle allows the system to be in such a combination of states, and the state AB + CD for the entire system is called an entangled state. While the product states AB and CD ascribes definite properties to BChls 1 and 2 (meaning, for example, that BChl 1 is excited and BChl 2 is in the ground state), the entangled state, since it constitutes a superposition, does not. The entangled state only says that there are possibilities concerning BChls 1 and 2 that are correlated, in the sense that if measurements are made, then if BChl 1 is excited then BChl 2 must be in the ground state; similarly if BChl 1 is in the ground state, then BChl 2 is excited. In other words, there is no way to characterize either one of them by itself without referring to the other as well. We can refer to each BChl alone when the two are in the product state AB or CD, but not when they are in the superposition AB + CD. It is the superposition of the two product states that produces the entanglement. This all can be generalized to a case of product states consisting of 2N BChl molecules, such as states (1). The superposition principle is employed in (2).

The state of $n$ qubits requires $2^n$ complex numbers to describe. Thus it seems qubits contain much more "information" than classical bits, since the state of $n$ classical bits is described by $n$ elements each either 0 or 1. It is impossible to retrieve all information from the state of $n$ qubits due to the fact that measuring the state destroys information. But when Nature evolves a closed quantum system of qubits, not performing any measurements, she does keep track off all the continuous variables describing the state. It seems that this hidden quantum information is what is harnessed through coherence and entanglement in LH-II rings of purple bacteria.

Both quantum Fourier transforms are instances of the hidden subgroup problem for the cyclic group $Z_N$ where N is 16 for *Rs. molischianum* and 18 for *Rps. acidophila*. In the hidden subgroup problem the task is to determine a generating set for H, where H is a subgroup of a group G. In our case the cyclic group $Z_N$ is the group G. Reference [27] illustrates the hidden subgroup problem for G = $(Z_N,+)$ (the additive group of integers mod N). The group G is mapped into the basis of the quantum states, that is, G = $\{|0>,|1>,...,|N-1>\}$, while H is mapped into H=$\{|0>,|d>,|2d>,...,|(M-1)d>\}$ where M is the order of the subgroup H and N = dM. M can be determined using the expression (29) from the reference [27]

$$\psi_f = \frac{1}{\sqrt{d}} \sum_{t=0}^{d-1} e^{\frac{2\pi i j_0 tM}{N}} |tM>$$

and the generating set for H is given by d = N/M. If M =1 then d = N and H=$\{|0>\}$. In this case the above expression is equivalent to (7) and this shows that the quantum Fourier transform for the group $Z_N$ is a hidden subgroup problem for a trivial subgroup H consisting of only neutral element which is mapped to the state $|0>$.

The idea that purple bacteria implement an efficient quantum circuit prompts interesting questions of qubit realization in LH-II of *Rs. molischianum*: Are four B850 BChls involved in realization of one qubit? Or, due to 8-fold symmetry, all BChls contribute equally to 4 qubits. How Haddamard $H$ and $R_p$ gates are built by BChls? Finding answers on these questions could open a possibility of using BChls as building elements in quantum computers.

**(5) Conclusions**

As pointed out by the physicist and the philosopher of science Percy Bridgman [28], positive convictions play an essential role in research: "Doubtless, there are many simple connections still to be discovered [in physics], and he who has a strong convinction of the existence of these simple connections is much more likely to find them than he who is not at all sure they are there".
The author of this paper has a strong conviction that purple bacteria implement an efficient quantum circuit for light harvesting. We have shown by suitable labeling of the lowest energy ($Q_y$) excited states of individual BChls, that the resulting exciton states are the quantum Fourier transform of the BChls excited states. For *Rs. molischianum* LH-II ring exciton states can be modeled as the four qubit quantum Fourier transform and we derived the explicit quantum circuit. Exciton states for *Rps. acidophila* LH-II ring cannot be modeled with an integer number of qubits.

A further research might reveal an efficient underlying quantum circuit for which the circuit shown in the Fig.1. is a reasonable approximation. Both quantum Fourier transforms are instances of the hidden subgroup problem. Future studies of LH-II rings in purple bacteria could open a possibility of using BChls as building elements in quantum computers. It is amazing that purple bacteria discovered efficient quantum circuits for light harvesting long time ago before we appeared.

**Acknowledgments**

The results presented on these pages are the outcome of independent research not supported by any institution or government grant.

**References**

[1] Borisov, A. Yu.; Freiberg, A. M.; Godik, V. I.; Rebane, K. K.; Timpmann, K. K. *Biochim. Biophys. Acta* **1985**, 807, 221.

[2] Sundstrom, V.; Van Grondelle, R.; Bergstrom, H; Akesson, E; Gillbro, T. *Biochim. Biophys. Acta* **1986**, 851, 431.

[3] Du, M.; Xie, X., Jia, Y.; Mets, L.; Fleming, G. R. *Chem. Phys. Lett*. **1993**, 210, 535.

[4] Hess, S.; Akesson, E.; Cogdell, R.; Pullerits, T.; Sundstrom, V. *Biophys. J.* **1995**, 69, 2211.

[5] Visser, H. M.; Somsen, O. J. G.; Van Mourik, F.; Lin, S.; Van Stokkum, I.; Van Grondelle, R. *Biophys. J*. **1995**, 69, 1083.

[6] Shreve, A. P.; Trautman, J. K.; Frank, H. A.; Owens, T. G.; Albrecht, A. C. *Biochim. Biophys. Acta* **1991**, 1051, 280.

[7] Timpmann, K.; Zhang, F. G.; Freiberg, A.; Sundstrom, V. *Biochim. Biophys. Acta* **1993**, 1183, 185.

[8] Kleinherenbrink, A. M.; Deinum, G.; Otte, S. C. M.; Hoff, A. J.; Amesz, J. *Biochim. Biophys. Acta* **1992**, 1099, 175.

[9] Hu, X. & Schulten, K. **(1997)** *Physics Today* 50, 28-34.

[10] J. Koepke, X. Hu, C. Muenke, K.Schulten, H. Michel, *Structure* **1996,** v.4, 581.

[11] McDermott, G., Prince, S. M.., Freer, A. A., Hawthornthwaite-Lawless, A. M., Papiz, M. Z., Cogdell, R. J., Isaacs, N. W., *Nature* **1995**, 374:517-521.

[12] P. W. Shor, "Polynomial-Time Algorithms for Prime Factorization and Discrete Logarithms on a Quantum Computer", Siam. *J. Comput*. 26 1484-1509 (**1997**), quantph/9508027.

[13] L. Grover, *Proceedings, 28th Annual ACM Symposium on the Theory of Computing (STOC)*, May **1996**, pages 212-219.

[14] S. Lloyd, "Universal Quantum Simulators", *Science*, 273, 23 Aug. **1996**.

[15] P. Benioff, *J. Stat. Phys*. 22, 563, **1980**.

[16] R. P. Feynman, "Simulating Physics with Computers", *International Journal of Theoretical Physics*, 21, Nos 6/7, **1982**.

[17] D. Deutch, "Quantum theory, the Church-Turing principle and the universal quantum computer", *Proc. R. Soc. Lond*. A 400, 97-117 **(1985)**.

[18] A. Steane, Rept. Prog. Phys. 61, 117-173 (1998).

[19] D. P. DiVincenzo, "Two-Bit Gates are Universal for Quantum Computation", *Phys. Rev. A* 51, 1015 **(1995)**.

[20] Lecture notes of J. Preskill at *http://www.caltech.edu/subpages/pmares.html*.

[21] D. S. Abrams, and S. Lloyd, "Simulations of many-body Fermi systems on a universal quantum computer", *Phys. Rev. Lett*. 79 **(1997)**.

[22] S. S. Somaroo, et al, "Quantum Simulations on a Quantum Computer", **quantph/9905045.**

[23] C. Zalka, "Simulating quantum systems on a quantum computer", *Proc. R. Soc. Lond*. A 454, 313-322 **(1998)**.

[24] M. A. Nielsen and I. L. Chuang, "Quantum Computation and Quantum Information", (Cambridge University Press, Cambridge, **2000**).

[25] N. E. Mavromatos, A. Mershin, D. V. Nanopoulos, "QED-Cavity model of microtubules implies dissipationless energy transfer and biological quantum teleportation", **arXiv:quant-ph/0204021** v1, 4 Apr 2002.

[26] Amir D. Aczel, "Entanglement The Greatest Mystery in Physics", (Four Walls Eight Windows, New York, **2002**).

[27] Chris Lomont, "The Hidden Subgroup Problem - Review and Open Problems", **arXiv: quant-ph/0411037** v1 4 Nov 2004.

[28] Glashow, Sheldon L. "The Charm of Physics", (Touchstone Books, Simon & Schuster, New York, **1991**), page 9.